# Frequency-doubled chirped-pulse dual-comb generation in the near-UV: Combined vs separated beam investigations of Rb atoms near 420 nm

JASPER R. STROUD,[1,*] AND DAVID F. PLUSQUELLIC[1]

[1]*National Institute of Standards and Technology, Boulder, CO 80305*
*\*jasper.stroud@nist.gov*

**Abstract:** We describe an electro-optic dual-comb system that operates in the near-infrared (near-IR) region to generate optical frequency combs in the near-UV by sum frequency generation in two configurations. The near-IR frequency combs are generated using chirped pulses that down convert the optical information into the radio frequency (RF) domain by a difference in the chirp bandwidths. Near (UV) combs at twice the near-IR bandwidth are obtained by sum frequency generation in a nonlinear crystal and detected by a hybrid photon counting detection system. We compare the results of studies of Rb near 420 nm using two optical arrangements where the near-IR combs are mixed in the crystal as combined or as separated beams. While the latter method enables phase retrievals, the combined beam method is superior for phase stability, power throughput for detection, and ease of alignment.  High order interleaving enables near-UV bandwidths near 4 cm$^{-1}$ for faint photonic sensing and spectroscopic applications.  The harmonic generation method is easily extendable across much of the titanium sapphire tuning range.

## 1. Introduction

The ultraviolet (UV) to visible region of the electromagnetic spectrum has provided vital information for key advances in diverse frontier areas that range from atom-based devices for optical frequency clocks [1], to Rydberg atom sensing of magnetic [2] and electric fields [3], to the emerging fields of quantum sensing [4,5] and imaging [6]. For an increasing number of these applications, optical frequency combs (OFC) provide easy access to the quantum states of interest and a robust route to the SI traceability of the spectroscopic measurements.

Mode-locked (ML) OFC methods have been transformational by providing high resolution spectroscopic signatures over multiple THz of optical bandwidth [7,8,9]. This technique has enabled many advances in trace gas detection in the near-infrared (IR) region and beyond [10,11]. The high peak power of the pulse train allows for efficient nonlinear frequency conversion across wide spectral regions from the microwave (MW) to vacuum UV (VUV) regions. However, for continuous wave (CW) generation in the UV region, the nonlinear mixing is primarily limited to frequency doubling in enhancement cavities in the 500 nm to 900 nm region [12,13,14]. For integrated photonic systems, sum frequency generation [15,16,17,18] has been demonstrated into the visible range using lithium niobate (LN) electro-optic (EO) modulators and ring resonators. However, generating OFCs at shorter wavelengths below 500 nm becomes increasingly difficult because of strong material dispersion and propagation losses. Moreover, for integrated LN devices, photorefractive effects from trapped charges have led to unacceptable phase drifts [19] although recent devices without this drawback have been reported across the visible region from 400 nm to 700 nm [20].

The region for interrogation of many atom-based devices is bounded to the tens of GHz range where the high power per comb tooth and the low relative phase noise of CW EO combs are well suited. Further, many other applications do not require THz of bandwidth provided by ML-OFCs, such as pressure sensing of a GHz wide spectra feature or tracking the position of a cavity resonance over tens of GHz for temperature or acceleration sensing [21,22,23]. Dual

EO-OFCs are often limited in optical bandwidth but make up for it in hardware simplicity [24,25]. Using either one or two non-phase locked optical sources, the mutual phase coherence of the two legs of an interferometer with EOs has been shown to provide excellent phase stability without complicated locking setups. These components also can be easily integrated for deployment of on-chip devices [26].

These attributes have made dual EO-OFC spectroscopy an invaluable tool in the infrared regions, but the lack of phase modulators in the UV region has made it difficult to directly implement this technique in that region. The UV region is readily reached using nonlinear mixing processes that double near-IR photons into the UV in a nonlinear crystal [27], but the conversion efficiency is too low for traditional detection. Even for ML-OFC methods, direct comb generation below 500 nm in the visible requires a comb-pumped nonlinear mixing process either with type 0, 1 or 2 mixing in nonlinear crystals [28,29] or for the VUV region, enhancement cavities together with noble gas jets [30,31] or other solid materials [32,33].

We have previously reported on using photon counting to accumulate interferograms for remote sensing in the near-IR region to achieve heterodyne detection at very low light levels [34]. In this article, we extend this technique into the near-UV region in combination with our dual comb differential chirped-pulse down-conversion technique [35,36,37,38]. A UV dual comb system has been recently reported that generates, in two stages, harmonic combs using a traditional dual EO-OFC method operating near 1.5 µm. In this work, we use a CW source near 840 nm and dual chirped pulses of an EO-OFC to generate UV combs in a single stage and in a single crystal.

Chirped pulse spectroscopy is now routinely used in microwave region [39,40,41], but is an emerging technique being applied as an optical frequency comb method [42]. Phase coherent repeated chirps generate flat OFC with comb spacings and spectral bandwidths determined solely by the programable chirp parameters [43]. This allows for high flexibility of the desired optical resolution, bandwidth, and recording time, depending on the sensing requirements of the system. We have recently described a dual comb technique to extend the bandwidth for spectroscopy far beyond the detection bandwidth of the system, allowing for an optical region of > 100 GHz region to be recorded within an RF detection bandwidth of < 500 MHz [35]. We now extend this technique in the near-UV region where molecular structure and reaction dynamics can be rapidly sampled at high-resolution.

The paper is structured as follows; section 2 describes the two experimental setups used to generate dual OFC in the near-UV. Section 3 describes the nonlinear mixing processes that up-convert the combs to the near-UV region, and data processing procedures applied to the differential chirp down conversion results. Section 4 illustrates the results for application of both methods to absorption measurements of a Rb reference cell near 420 nm. Finally, section 5 discusses the benefits and drawbacks of each system design.

## 2. Experiment

### 2.1 Combined beam (CB) and separate beam (SB) system configurations

The experimental setup used to generate dual OFCs is shown in Fig. 1(a). The system is similar to that described to generate frequency combs in the THz region based on difference frequency generation [37]. As mentioned in this work, this same approach is applied here to instead generate UV combs via sum frequency generation. Briefly, a cavity stabilized Ti:Sapp ring laser (720 nm to 980 nm) pumped with 10 W of 532 nm light generates more than 250 mW of near-IR radiation. The laser is externally locked to a reference cavity to achieve a frequency stability of < 1 MHz in short term (< 10 sec). A small portion is also split off by an optical wedge for wavemeter readout and for long-term drift control (< 2 MHz) using a HeNe stabilized reference cavity [44,45].

The remaining output is free space coupled into a polarization maintaining single mode fiber (PMSMF) with > 60 % efficiency and split into the two legs of the interferometer that define the signal (SIG) and local oscillator (LO) combs. Each leg consists of an acousto-optic modulator (AOM, Brimrose, TEM-50-2-60-850-2FP, (5-6) dB insertion loss) followed by an electro-optic phase modulator (EOM, EO-space, PM-5SE-10-PFA-PFA-850-LV, 3 dB insertion loss). AOM1 and AOM2 are driven with function generators (FG, Keithley, 3390, amplified to 1 W) at 50.000 MHz and 49.996 MHz, respectively, to generate a 4 kHz beat note that separates the positive and negative EOM sidebands. The EOMs are driven by chirped pulses from an arbitrary waveform generator (AWG, Keysight M8195, 65 GS/s, 8-bit), each amplified by up to 1 W over the bandwidth from 1 GHz to 15 GHz. The overall optical insertion loss for this system is ≈ 10 dB.

The chirped pulse waveforms define the OFC properties where the chirp duration is inversely related to the comb tooth spacing, the chirped interval defines the comb bandwidth, and the drive voltage defines the Bessel function distribution of EOM sideband amplitudes [35]. The two chirped pulses in the SIG and LO legs are defined to have the same chirp duration but different chirp bandwidths. The bandwidth difference defines the down-converted bandwidth in the radio frequency domain, $\Delta f_{RF}$,

$$\Delta f_{RF} = \Delta f_{LO} - \Delta f_{SIG} \tag{1}$$

where $\Delta f_{SIG}$ and $\Delta f_{LO}$ are the SIG and LO bandwidths, respectively. The different bandwidths also result in different chirp rates,

$$\alpha_{SIG/LO} = \frac{\Delta f_{SIG/LO}}{\tau_{CP}} \tag{2}$$

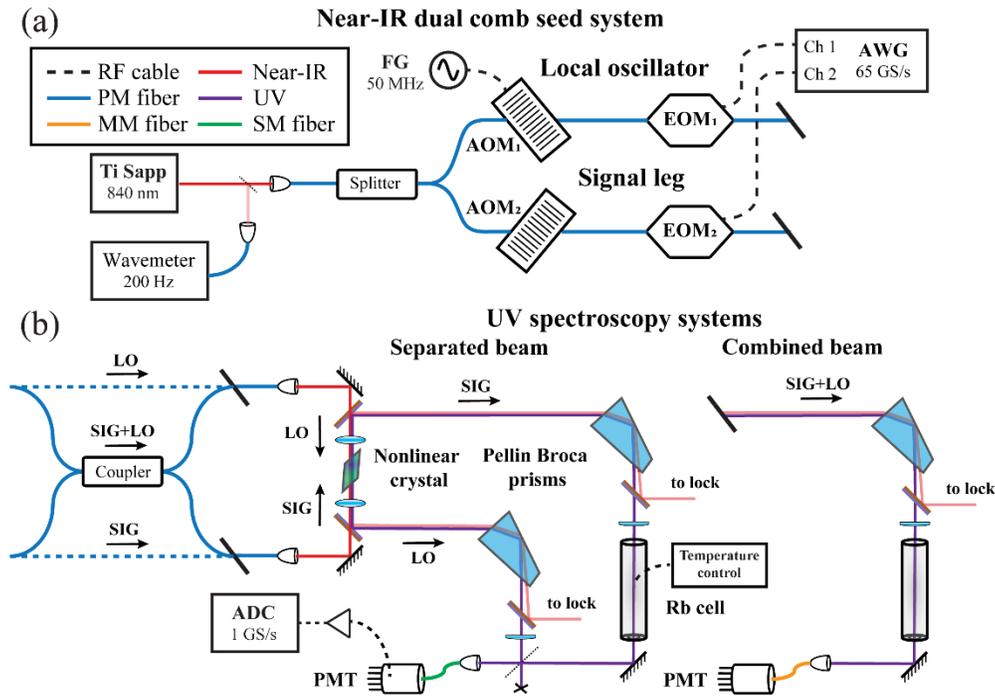

Fig. 1. (a) The common elements of the dual comb seed system. (b) UV generation for the CB and SB configurations showing the nonlinear mixing products are dispersivly filtered using Pellin Broca prisms before sampling the Rb reference cell then detected by a PMT.

For this work, the SIG chirp is defined from $f_{SIGstart} = 1$ GHz to $f_{SIGstop} = 15$ GHz and $\tau_{CP} = 50$ μs duration. The LO chirp is defined from $f_{LOstart} = f_{SIGstart} - f_{RFstart}$ to $f_{LOstop} = f_{SIGstop} - f_{RFstop}$, where $f_{RFstart} = 3$ MHz and $f_{RFstop} = 5$ MHz. Therefore, the system down converts 14 GHz of optical bandwidth in each sideband to $\Delta f_{RF} = 2$ MHz at the detector.

Two different arrangements are explored to generate near-UV combs near 420 nm by Type I phase matching inside a Brewster-cut lithium triborate (LBO) crystal (10 mm long, φ=27.9º, θ=90º for P-polarized 840 nm light). In the first arrangement shown in Fig. 1(b) referred as the combined beam (CB) method, the two near-IR legs are first combined in a 50/50 fiber coupler and one of output legs is collimated and focused inside the LBO crystal with a f = 5 cm achromatic lens (f=focal length). The sum frequency UV output having S-polarization is collimated with a second f = 5 cm lens, polarization rotated with a waveplate and spatially filtered from the residual fundamental following a Pellin Broca prism. The SIG+LO UV beams pass through f = 30 cm cylindrical lens to correct astigmatism of the Brewster-cut crystal and then through a heated (113 C) 7.5 cm long Rb reference cell equipped with tilted windows. The output beam is coupled into a multimode fiber (MMF, 50 μm core) for detection on a low-dark-count photomultiplier tube (PMT, EMI 9813QA, -1800 V, quantum efficiency of ≈ 25 % from 190 nm to 450 nm). The typical near-IR power at the crystal is 5 mW to 10 mW and while too low to be measured on a power meter, the conversion efficiency to the UV is expected to be near 0.01 % [46,47].

In the second arrangement, also shown in Fig. 1(b) and referred as the separated beam (SB) method, the SIG and LO UV beams are generated separately in a crossed path configuration in a single LBO crystal where the SIG beam travels in one direction through the crystal, while the LO beam travels in the opposite direction. The arrangement simplifies the tuning of the system by the simultaneous Type I angle phase matching. This type of SB method has been discussed previously although doubling was performed in two crystals. A pair of dichroic mirrors (Thorlabs, DMLP505) on either side of the crystal are used to reflect the UV light (R > 99 %) and pass the near-IR (T > 90 %) to the crystal. Each leg is passed through a half waveplate and Pellin Broca prism to separate the residual near-IR. The SIG is sent through the Rb reference cell before being overlapped on a 50/50 beam combiner with the LO. Only one of the outputs from the splitter is coupled into a single mode fiber (SMF, Thorlabs, P5-305AR-2, 4 um core) and detected with the PMT system. The second output is currently discarded although a second PMT system could be used to enhance the signal-to-noise ratio (SNR) since the same information is contained in both halves except for a simple 180º phase shift. It is further noted that for the CB method, the discarded output from the final near-IR fiber combiner in Fig. 1(b) could also be frequency summed in the crossed beam configuration and simultaneously passed through the Rb cell to improve the SNR again or to serve as the reference channel in a dual beam arrangement (not discussed here).

For both the CB and SB methods, phase locking of the 4 kHz beat note in the near-IR was performed for coherent averaging over periods of up to 10 sec using a scheme described elsewhere [35]. To achieve the best phase lock stability in the UV, the two residual near-IR beams reflected from the dichroic mirrors in Fig. 1(b) are combined on a polarization beam splitter, coupled into a 10 μm fiber and detected using a 200 MHz photodiode. The modulation depths (MDs) of the 4 kHz beat note typically ranged from 30 % to 40 %. For the CB method, the near perfect overlap in the PMSMF of the co-propagated SIG and LO beams resulted in a MD that was always near 100 %.

For both methods, the PMT output(s) is first amplified by a high speed ($\tau \approx 1$ ns) low noise transimpedance amplifier (LNA, Femto, HCA-400M-5K-C) and digitized at 1 gigasample/s (GS/s) in an analog-to-digital converter (ADC, Gage, CSE123G2, 12-bit) equipped with a field programmable gate array for data streaming to computer memory. For up to 2 channels, photon counting and photocurrent averaging are performed on the data streams, and the accumulated counts and average current records are stored to disk with a throughput of > 80 % to enhance the dynamic range and to eliminate issues with pulse pileup error [35]. As we first described for a multi-heterodyne remote sensing system near 1.6 μm [34], interferograms are formed by the accumulated counts as well as by the averaged photocurrent. Typical signal levels in this work range from 10 million counts/s (CPS) to 50 million CPS (dark count rates are < 250 CPS), the higher of which is the upper threshold where the pulse pile-up error begins to appear. Because of these high count rates, the interferograms and comb spectra from the count and current records are essentially identical therefore used interchangeably in this work.

The AWG and ADC systems are triggered at 800 Hz by a pulse delay generator (SRS, Model 535) where each interferogram consists of a 1 ms-long 1 megasample (MS) record for each of the counts and current channels. For both methods, the phase coherence is sufficiently long to coadd 4,400 interferograms during 5.5 sec in real time at 80 % throughput, prior to storage on a hard disk. Typically, measurements were performed over a 15 min period where the final comb spectra were obtained from the average of the Fourier transforms (FTs) of each record (either counts or current). To remove standing wave effects arising from the Rb cell windows, background spectra are obtained by detuning the laser by (1-2) cm$^{-1}$. The AWG, ADC and FGs are all disciplined using the same 10 MHz Rubidium clock reference.

### 3. Near-UV comb generation

#### *3.1 Nonlinear mixing of dual chirped pulses*

The near-UV interferograms averaged over 15 min periods are shown in Fig. 2(a) for the SB (blue, top panel) and CB (red, bottom panel) methods. For the CB method, the 4 kHz beat note (i.e., four repeated patterns) is readily apparent, while for the SB case, two interleaved oscillations are observed at 8 kHz. Figure 2(b) shows an expanded portion of the interferogram in Fig. 2(a) near where a change in frequency occurs from the end of one chirp to the start of the next. The additional features in the CB interferogram relative to the SB response are due to the extra mixing products that are produced in the nonlinear conversion in the crystal.

For the CB nonlinear mixing process, the first-order mixing products result from the six different waves that pass through the crystal. These input waves can be described at any given time by,

$$E_{input}(t,\omega) = E_{SIG}(\omega \pm \omega_{SIG}(t)) + E_{LO}(\omega \pm \omega_{LO}(t) + \Delta\omega_{AOM}) + E_{SIG_0}(\omega) + E_{LO_0}(\omega + \Delta\omega_{AOM}) \quad (3)$$

where $E_{SIG}(\omega \pm \omega_{SIG}(t))$ are the two chirped sidebands generated by the SIG leg, $E_{LO}(\omega \pm \omega_{LO}(t) + \Delta\omega_{AOM})$ are the corresponding LO chirped sidebands, and $E_{SIG_0}(\omega)$ and $E_{LO_0}(\omega + \Delta\omega_{AOM})$ are the two residual carrier waves. The terms $\omega_{SIG}(t)$ and $\omega_{LO}(t)$ describe the chirps of the SIG and LO, respectively. The beat note from the difference in the frequency shifts of the two AOMs is added to the LO terms as $\Delta\omega_{AOM}$. When this near-IR light is incident on a detector, the mixing products include RF combs of the positive and negative sidebands that are separated by the AOM beat note frequency,

$$LPF\{E_{input}(t,\omega)^2\} \ni E_{SIG}E_{LO}(\omega_{RF}(t) - \Delta\omega_{AOM}) + E_{SIG}E_{LO}(-\omega_{RF}(t) + \Delta\omega_{AOM}) \quad (4)$$

where $LPF$ is a low pass filter response and $\omega_{RF}(t) = \omega_{SIG}(t) - \omega_{LO}(t)$ is the down-converted difference frequency product sampled by the square law detector that maps the spectral information in the optical region to the RF region. The first and second terms in Eq. 4 are the positive and negative sideband combs, respectively.

In the LBO crystal, the sum frequency mixing process is modeled by a high pass filter of the squared input waves in Eq. 3 that produces many different combs corresponding to the different sum frequency wave interactions,

$$E_{crystal}(t, 2\omega) \ni HPF\{E_{input}^2(t, \omega)\} \tag{5}$$

The output of the nonlinear crystal will generate an array of combs that must map uniquely to reveal the underlying spectroscopy. There is a special class of interactions defined as second harmonic generation, where the two input waves in the nonlinear interaction are the same frequency. The few second harmonic generated terms are,

$$E_{crystal}(t, 2\omega) \ni E_{SIG}^2(2\omega \pm 2\omega_{SIG}(t)) + E_{LO}^2(2\omega \pm 2\omega_{LO}(t) + 2\Delta\omega_{AOM}) + DC \tag{6}$$

where key to this unique mapping is the LO term that now has an AOM shift twice that of the fundamental comb described in Eq. 3, with the DC term referring to the square of the unchirped terms. When these terms are incident on a photodetector, the square law RF output is proportional to,

$$LPF\{E_{crystal}^2(t, 2\omega)\} \ni E_{SIG}^2 E_{LO}^2(2\omega_{RF}(t) - 2\Delta\omega_{AOM}) + E_{SIG}^2 E_{LO}^2(-2\omega_{RF}(t) + 2\Delta\omega_{AOM}) \tag{7}$$

Eq. 7 is nearly identical to Eq. 4, except the comb has twice the optical and RF bandwidth (and twice the number of comb lines). Further, these combs have twice the AOM shift so they are spectrally separated from the fundamental comb. Figure 2(c) shows the different orders of the comb spectra from the CB system in blue, and the SB system in red (offset for clarity). The second-order combs described in Eq. 7 are shown from 6 MHz to 10 MHz in Fig 2(c).

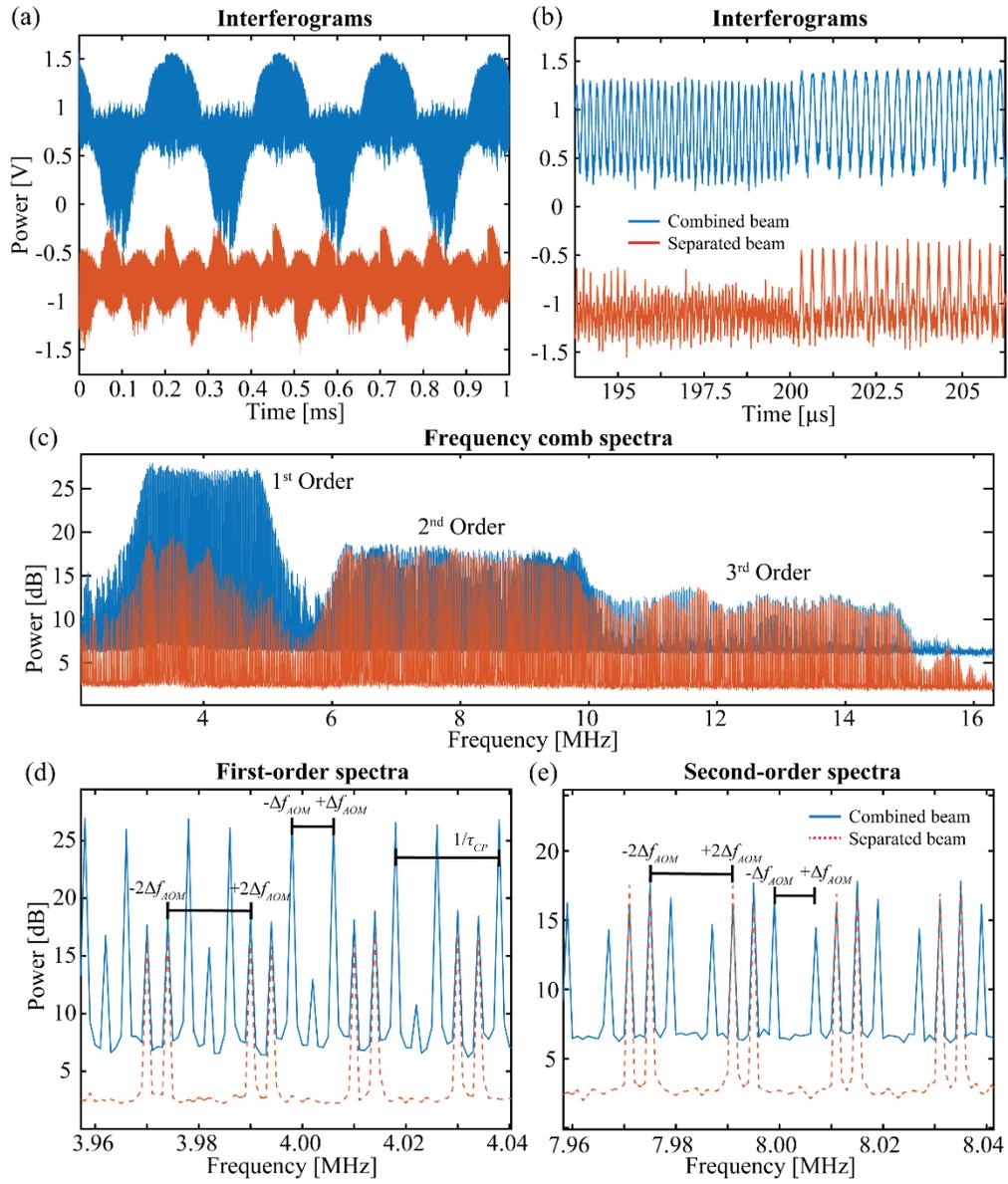

Fig. 2. (a) The interferogram of the CB system in blue and SB system in red, with (b) the zoomed in section showing the transition between chirped pulses. (c) The comb spectrum, with (d) the first-order comb lines and (e) the second-order comb lines. The corresponding SB data containing only frequency doubled comb lines is shown in red and offset for clarity.

Figures 2(d) and 2(e) show expanded portions of the first-order and second-order spectra, respectively, where lines appear separated by twice the AOM beat frequency in both configurations.

For both methods, there are two other combs generated by the sum frequency generation process that form a comb at twice the AOM beat note separation. The products between the chirp and its corresponding carrier are,

$$E_{crystal}(t, 2\omega) \ni E_{SIG}E_{SIG_0}(2\omega \pm \omega_{SIG}(t)) + E_{LO}E_{LO_0}(2\omega \pm \omega_{LO}(t) + 2\Delta\omega_{AOM}) \quad (8)$$

As before, these terms mix on the detector to produce a comb that spans the same amount of bandwidth as the fundamental comb but have a unique mapping in the near-UV spectrum due to the AOM beat note doubling of the carrier.

$$LPF\{E^2_{crystal}(t, 2\omega)\} \ni E_{SIG}E_{SIG_0}E_{LO}E_{LO_0}(\pm\omega_{RF}(t) \mp 2\Delta\omega_{AOM}) \quad (9)$$

This represents a mixed order comb, generated from the product of the first-order chirp and carrier wave. Figure 2(d) shows the first-order comb, where the comb teeth are at twice the AOM frequency difference for both system designs. It follows that any combination of higher EOM orders will mix to generate UV combs. The drive voltage can be easily adjusted in the AWG to optimize for higher orders at drive voltages above $V_\pi$ (≈ 3 V) of the modulators. The distribution of near-UV comb amplitudes is proportional to the square of the product of the Bessel function amplitudes. The third-order comb shown in Fig. 2(c) spans from 9 MHz to 15 MHz and is produced from the product between the first- and second-order fundamental combs.

The second harmonic and the inter-order mixing products describe all the comb spectra generated by the SB system and are shown in red in Figs. 2(c), 2(d) and 2(e). In contrast, the results for the CB system (in blue) clearly have several additional combs. These added features are due to the mixing between the SIG and LO chirps that generate UV combs but with only a single AOM beat note shift. Further, these products form combs that are not unique in the RF domain because multiple optical spectral regions are mapped onto the same RF comb. (In the supplemental materials, we illustrate how the degenerate combs map the fringes of a high finesse etalon at 420 nm.) However, as shown in Fig 2(d) and 2(e), the degenerate combs are cleanly separated from the uniquely mapped ones because of the differences in the beat frequencies. While both methods produce uniquely mapped spectra, the SB method produces a far cleaner spectrum and allows for more possible combinations of beat note differences and interleaved EOM orders for the generation of different spectral resolutions and bandwidths. As a case in point, the degenerate comb begins to overlap with the unique fifth-order comb with the system settings used in this work (4 kHz beat note, 50 µs chirp duration).

We also find that at comparable count rates, the SNR of the SB comb with far fewer comb lines shows a lower-than-expected improvement relative to the CB comb. While the SB interferogram in Fig. 2(c) was acquired at a 35 % higher count rate (15 MCPS vs 11 MCPS for the CB method), the SB comb spectra in Figs. 2(d) and 2(e) show only between 3 dB to 5 dB higher SNRs, respectively, which are not enough to account for the very strong degenerate combs shown in Fig. 2(c). Part of this is rationalized because of the low crystal mixing efficiency in single pass ($\approx$ 0.01 %) where the absence of pump wave depletion leads to fixed powers in each unique comb. Further, the integrated power in the DC component of the interferograms from Eq. 6 is found to be reduced by more than 2-fold for the CB method where the degenerate combs must receive some power to account for the similar count rates observed for the two methods.

*3.2 Data processing*

The down-converted optical frequency comb spectra shown in Figs. 2(c), 2(d) and 2(e) are obtained from the 15 min average of the Fourier transforms of multiple records like that shown in Fig. 2(a). The comb teeth obtained for the Rb spectrum near 420 nm are normalized against the background comb to produce the frequency domain spectrum. Prior to normalization, the selected comb teeth of a given order are inverse Fourier transformed then normalized to generate the corresponding time domain spectrum. Due to the quadratic phase imparted onto the system response by the chirped LO pulse, the transient response is magnified in the frequency domain [35]. A further complication occurs for the CB method where both the SIG and LO sample the Rb spectrum and then act to magnify each other, resulting in two different magnification factors of opposite sign (i.e., direction of ripples),

$$m_{SIG/LO} = \frac{\Delta f_{LO/SIG}}{\Delta f_{SIG/LO} - \Delta f_{LO/SIG}} = \pm \frac{\Delta f_{LO/SIG}}{\Delta f_{RF}} \qquad (10)$$

where $m_{SIG}$ and $m_{LO}$ are the magnification factors of the SIG and LO legs, respectively. The frequency domain spectra experience a magnified chirp rate, $\alpha_{f_{SIG/LO}}$, where the temporal response is $m_{SIG/LO}$ times faster than the natural chirp rate, $\alpha_{SIG/LO}$,

$$\alpha_{f_{SIG/LO}} = m_{SIG/LO} \alpha_{SIG/LO} \qquad (11)$$

The inverse Fourier transform removes the quadratic phase response imparted on the frequency domain line shape to give the unmagnified (natural) spectrum. When the chirp rate in the frequency domains exceeds the response time of the system (which is proportional to the spectral width), the rapid passage effects will cause oscillations (ripples) that distort the symmetric line shape (oscillations may even be observed in the time domain for even faster chirps that were not performed here) [37]. Figures 3(b) and 4(b) show the SB and CB model predictions of a single resonance in the frequency domain where these ripples are observed in one direction only or in both directions, respectively. For the SB method, the single-ended magnification arises from the LO quadratic phase alone in contrast to the CB method where complimentary magnifications from Eq. 11 occur in both directions by the simultaneous passage of the SIG and LO through the sample [25].

For both methods, the unmagnified time domain spectrum shown in Fig 3(a) and Fig 4(a) are recovered after inverse Fourier transformation. The SB method samples the true magnitude and phase of the sample in both the time and frequency domains. The CB method samples the intensity (squared magnitude) spectrum, while in the frequency domain, magnification differences in the two directions creates a non-zero phase response that is the difference of the magnified LO and SIG phase spectra.

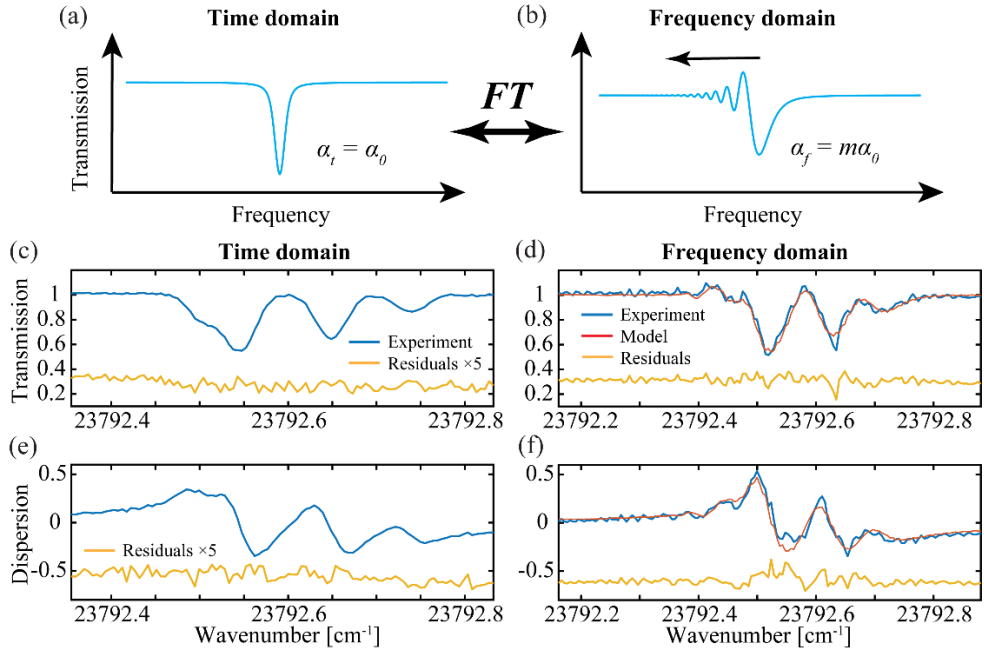

Fig. 3. (a) Time domain resonance shows a symmetric line shape while (b) the frequency domain data shows rapid passage in one direction. (c) Time domain amplitude and (e) phase spectra in blue, and the residuals (x5) in yellow. The frequency domain (d) amplitude and (f) phase spectra in blue, the models in red and unscaled residuals in yellow.

## 4. Results

### *4.1 Separated beam (SB) results and analysis*

We apply the SB method to measure the hyperfine absorption features of Rubidium (Rb) atoms near 420 nm in a 7.5 cm long cell that is insolated and heated to 113 C with heating elements on the cell and on each tilted window. The laser is locked at 11896.3523 cm$^{-1}$ to center the near-UV comb near the Rb multiplet at 713.285 THz. The comb samples across the main four-line structure of Rb where each line consists of six hyperfine components [48]. The spectra span from 2 GHz to 30 GHz on each sideband (where the time axis is converted to frequency using the SIG chirp rate). The SIG magnification factor for this data is $m_{SIG}$ = 7499 from Eq. 10 for an optical LO bandwidth of 29.996 GHz and a RF bandwidth of 4 MHz. For the LO, $m_{LO} = 7500$ because the SIG optical bandwidth of 30 GHz.

With careful alignment of the two beams into the single mode fiber, we obtained about 15 million CPS at the PMT and nearly equal contributions from the SIG and LO legs. The second-order time and frequency domain amplitude spectra are shown in blue in Figs. 3(c) and 3(d), respectively, and the corresponding phase spectra are shown in Figs. 3(e) and 3(f). The fractional absorption of the amplitude transmission spectrum shown in blue is nearly 50% and the rapid passage effects in the frequency domain are seen to extend out in one direction only as predicted in Fig. 3(b). The spectrum is fit by floating the amplitudes and positions of four lines (see Table I below) using the measured temperature to fix the Gaussian line shape widths and vapor pressure of Rb atoms in the cell (the pressure broadening Lorentzian contribution is negligible). The scaled (x5) residuals from the fits of the amplitude and phase spectra in the time domain show no clear structure above the noise and both have root-mean-square (RMS) standard deviations of < 0.9 % at the bottom of Fig. 3(b).

The best-fit time domain line shape in Fig. 3(c) is then used to predict the observed frequency domain profiles shown in Figs. 3(d) and 3(f). The magnified responses are shown in red in Figs. 3(d) and 3(f). To generate these profiles the simulated line shape is rescaled to higher temporal and spectral resolution to capture both the bandwidth of the chirp, as well as the resolution of the down converted RF signal. The simulated sample spectra are mixed with the SIG chirped waveform in the frequency domain and is then down-converted by mixing with the LO in the time domain (see supplemental materials for a full discussion). The fits are in reasonable overall agreement with the observed spectra although the residuals are now shown unscaled.

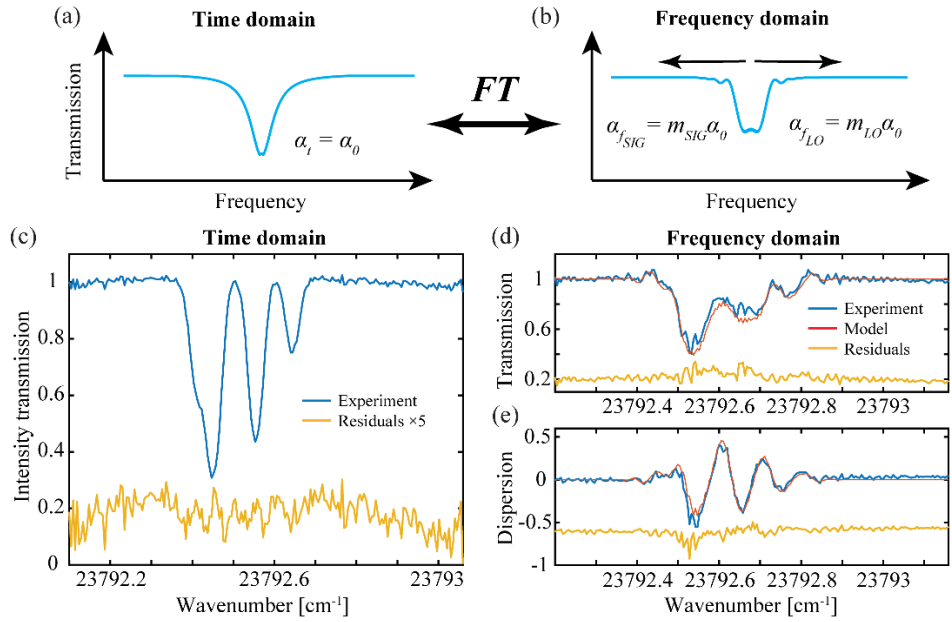

Fig. 4. Model predictions for the CB system that give (a) the intensity transmission spectrum in the time domain, and (b) the frequency domain spectrum that show rapid passage effects in both directions. (c) The experimental intensity transmission spectrum in the time domain (blue) and the magnified residuals (x5, yellow). (d) The observed (blue) and predicted (red) frequency domain transmission spectrum and (e) dispersion spectrum (red) shown with unscaled residuals (yellow). See text for details.

*4.2 Combined beam (CB) results and analysis*

With minimal change to the experimental alignment, we switched over to the CB method by changing a few fibers and by removing the final SIG/LO beam combiner in Fig. 1(b). For these measurements, the laser frequency was locked to 11896.4835 cm$^{-1}$. To make comparisons with the SB results, we attenuated the count rate on the PMT to about 11 million CPS. The second-order time and frequency domain spectra are shown in Fig. 4. The fractional absorption of the intensity (squared magnitude) transmission spectrum shown in blue in Fig. 4(c) now exceeds 70% in the time domain. The spectrum is fit by floating the intensities and positions of four lines. Like the SB results, the scaled (x5) residuals indicate no apparent residual line shape errors that exceed the noise. The RMS standard deviation of the residuals is < 1 % in the time domain.

The best-fit time domain profile in Fig. 4(c) is then used to predict the observed frequency domain profiles shown in Figs. 4(d) and 4(e). Unlike the SB method, this line shape is first mixed with both the SIG and LO chirped waveforms in the frequency domain and then mixed with the complimentary waveform in the time domain (see supplemental materials). This results in the predicted temporally magnified line shape functions shown in red in Figs. 4(d) and 4(e) with the residuals shown unscaled.

Interestingly, for both CB and SB data, the apparent SNR of the frequency domain is much worse than in the time domain. In contrast to other studies of ours where this same procedure was applied, the residuals in Figs. 3(d) and 3(f) and Figs. 4(d) and 4(e) show that the forward transformation of the best-fit time domain data does not fully model the observed responses in the frequency domain which suggests the higher frequency structure in the residuals is due to magnified spectral features. Additional spectroscopic information appears to be needed to properly account for this structure. Although not explored further here, the additional features may be associated with the unresolved hyperfine structure that make up each of the four main features [48].

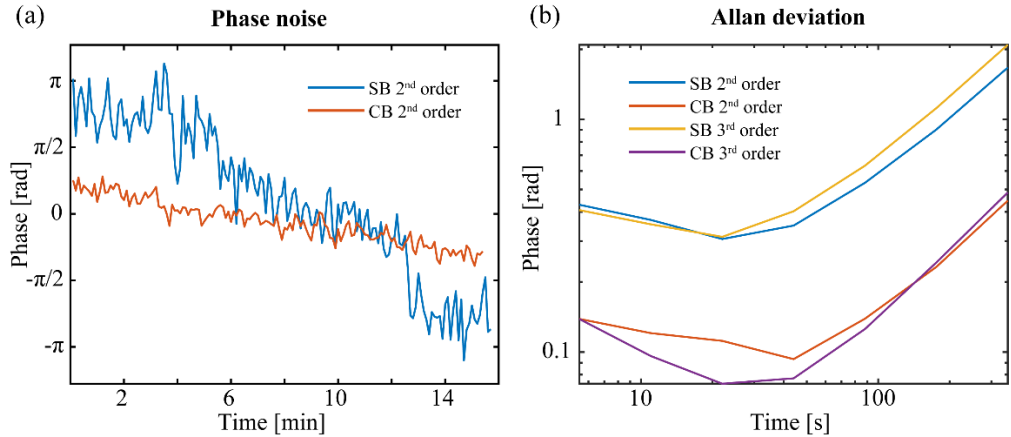

Fig. 5. The average phase angle of the second order comb teeth as a function of time for the SB method, in blue and CB case, in red. (b) The Allan deviation plots of the four data sets, showing the significant short- and long-term improvement in the phase stability using the CB system.

*4.3 Comparison of the phase instabilities of the SB and CB methods*

The phase instability of the two methods was evaluated using the raw data records acquired on 5.5 s intervals which corresponds to 4.4 s records at 80 % throughput. To illustrate the actual phase drift in the UV interferogram, the average phase angle of the comb teeth (8 kHz beat note) in the second and third orders are compared. The results are shown as a linear function of time for both methods in Fig. 5(a). It is clear from the SB results shown in blue that this configuration suffers from long-term phase drift approaching $2\pi$ rad over the 15 min interval relative to a drift near 2 rad for the CB method, a nearly 3-fold improvement. The Allan deviation analysis shown in Fig. 5(b) substantiates this conclusion and additionally shows that the short-term phase jitter of the CB method is also ≈ 3-fold better than that of the SB method. The turning point in the long term drift rate of the CB method suggests that signal averaging would be possible over 60 sec intervals without a significant degradation of the MD.

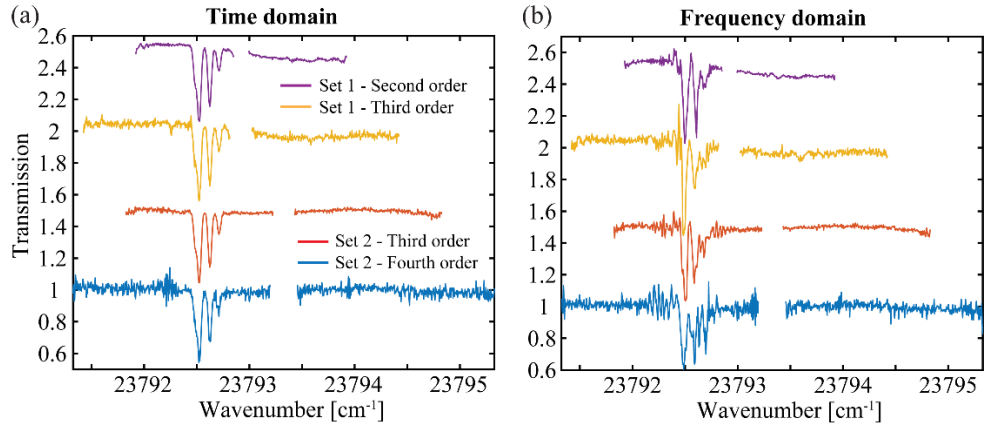

Fig. 6. (a) The time domain amplitude spectra and (b) frequency domain spectra of two data sets showing the bandwidth expansion form second order to the fourth order.

*4.4 Bandwidth expansion*

The drive voltage on the AWG is increased to generate higher-order outputs from the EOMs in the near-IR and as discussed earlier, for the higher-order mixed combs in the UV. This control allows for a trivial way to expand the bandwidth of the system. Figure 6 shows the time and frequency domain amplitude transmission spectra of two sets of data acquired with the SB method. For the first data set, the second- and third-order amplitude transmission spectra are shown as the purple and orange traces in Fig. 6 and span 2 cm$^{-1}$ and 3 cm$^{-1}$, respectively (the second-order spectrum is the same as in Fig. 4). For the second data set, the drive voltage is increased by ≈ 30 %, and the laser frequency is adjusted to 11896.5548 cm$^{-1}$ to better center the Rb profile in the negative sideband region. The third-order and fourth-order spectra are shown as the red and blue traces in the lower sections of Fig. 6, respectively. First notice that the width of the center gap region increases with the bandwidth of the order. Further, since the chirp rates and therefore, the magnifications in the frequency domain depend on order, the magnitudes of the ripples are seen to increase with order in Fig 6(b). However, these effects are completely absent in the back transformed spectra shown in Fig. 6(a). While the bandwidth coverage of the second-order comb is more than sufficient to measure the Rb absorption feature at 420 nm, the fourth-order comb enables increased spectral coverage of nearly 4 cm$^{-1}$ (in blue). We have also previously demonstrated the fourth-order comb generation in the near-IR which for this method, would translate to an eighth-order UV comb [35]. While this work is mainly centered in the near-UV region, we expect a similar performance down to 360 nm with a simple change of the optics set in the Ti:Sapp laser and with the substitution of a second LBO crystal that is centered at 770 nm.

The uncertainties of the fit parameters were estimated from the series of fits across several data sets. The data sets included the two second-order spectra discussed above for the SB and CB methods and two additional fits of the third-order spectra obtained at the higher drive voltage in Figs. 6(a). The average parameters and uncertainties (Type B, k=1 or 1σ) of the line centers and intensities of each of the four features in the Rb spectrum are given in Table 1. The precision of the line centers is < ±20 MHz for the $^{87}$Rb transitions and < ±l0 MHz for the $^{85}$Rb transitions. The absolute frequencies are estimated to be within ±100 MHz based on the wavemeter reading following calibration with two polarization stabilized HeNe lasers (±50

MHz). The precision of the line intensities are < 4 %. The uncertainties do not include the propagated errors from uncertainties in the cell temperature and cell path length although we estimate changes to be < ±1 % over the period of these measurements.

Table 1. **Best-fit line frequencies and intensities of the $6P_{3/2} \leftarrow 5S_{1/2}$ transitions of the $^{87}$Rb and $^{85}$Rb isotopologues. Uncertainties shown in the least significant digits are determined from the RMS differences (Type B, k=1 or 1σ) in four independent spectral fits of the four main-line features (see text for details).**

| Ground state | Frequency (THz) | Intensity (cm$^{-1}$/mol cm$^{-2}$ ×10$^{-16}$) |
|---|---|---|
| $^{87}$Rb F″ = 2 | 713.281618(15) | 3.07 (3) |
| $^{85}$Rb F″ = 3 | 713.282791(7) | 7.18(28) |
| $^{85}$Rb F″ = 2 | 713.285883(7) | 5.25(12) |
| $^{87}$Rb F″ = 1 | 713.288586(19) | 1.80(7) |

## 5. Discussion

The advantages and disadvantages of the combined beam (CB) and separated beam (SB) methods demonstrated in this work are now enumerated to help guide which approach is best suited for a particular application. One area of particular interest is the ease of which beam overlap can be achieved to ensure the highest possible MD at the detector. For a given count rate, the relative magnitude of the MD translates directly to the SNR of the comb lines since anything less than 100 % just shifts the optical amplitude to the DC component of the Fourier transform (see Eq. 11). As we demonstrate here for the CB method, one way to achieve near 100 % MD at the detector is to use the CB output from a PM fiber. Near perfect overlap at the detector was achieved even after frequency up-conversion in the crystal and the fiber coupling to the detector using a large 50 µm-core MMF. The high efficiency of MMF coupling led to PMT saturation with signal levels exceeding > 250 million CPS (i.e., the maximum count rate possible) and therefore, significant attenuation was required using neutral density filters to reduce rates for linear detection (10 million CPS to 50 million CPS). To achieve nearly 100 % MD at the detector using the SB method required careful collimation of both beams into a 4 um-core SMF. Further, the efficiency of SMF coupling of the two beams was strongly dependent on the matched wavefronts, focus, and the degree of spatial mode overlap of the coupled beam components. Since these conditions could only be partially achieved by compensation of astigmatic UV beams using the cylindrical lens, the maximum count rate achieved for SM coupling of both beams was 20 MCPS. Initial attempts at SB coupling through 300 µm and 50 µm MMF also led to PMT saturation but at unacceptable low MDs of < 2 % and < 25 %, respectively.

A second important issue that differs for the two methods is the degree of phase instability since phase jitter at the detector over the averaging period results in a shift of optical amplitude to the DC component of the RF comb spectra. Hence, the impact of phase error is similar to the problem of imperfect beam overlap in MMF that reduces the MD at the detector. As discussed above in Fig. 5, the phase instability of the CB method was shown to be ≈ 3-fold smaller than the SB method. For the CB setup, instabilities primarily arise from relative phase drift in the fibers between the SIG and LO legs of the interferometer (see Fig. 1(a)). The thermal and acoustic isolation of these fiber components led to a small and slow drift that was readily corrected for by the phase lock system [35]. The beat note signal for the lock is generated from the residual near-IR combined beam after up-conversion which, for ease of alignment, is coupled to the detector using a 2 m long MMF with a 10 µm core. As in the UV, the MD in the near-IR was near 100 %. We note that the faint UV signals from the PMT precluded their use in an active feedback loop for phase corrections. For the SB method, the interferometer over which phase errors can occur is extended to include the free-space paths in the UV of the SIG and LO legs. Despite the stable single-crystal crossed-beam scheme used for UV generation, only after the UV beam path was fully covered were phase errors small enough not to wash out the spectrum. Further, the sample itself can cause large phase fluctuations, where, for example, in an initial attempt, a temperature gradient between the windows and body of the Rb cell caused complete washout of the SB interferogram while the CB interferogram remained phase stable. Finally, like the UV beam alignment, the SB method required the careful collimation and overlap of the residual near-IR beams into the MM fiber, which further degraded the phase lock because of the reduced MD (< 50 %) at the detector. A third advantage of the CB system is in the simple implementation of a dual beam system where, prior to the sample, a split off portion of the UV beam would probe an empty reference path for normalization. The dual beam method is a requirement for spectra where detuning to an off-resonance region is not feasible as we will report elsewhere. For example, the absorption spectrum of $NO_2$ roughly extends from 350 nm to 450 nm, a span that is far beyond the tuning range of our current LBO crystal, making off-resonance tuning impossible. A similar dual beam approach for the SB method would require matching the path lengths for temporal pulse overlap of the SIG and LO legs in both the sample and reference arms.

The SB method has a few distinct advantages over the CB method that may be of importance in certain specialized applications. Because the SIG and LO beams are combined after the sample, both the amplitude and phase spectra of a sample are recovered with similar SNR as shown in Figs. 3(c) and 3(e). These independent measurements give insight into the complex behavior of light interactions in a sample and, because of the narrower line widths in the amplitude spectrum, can enhance the resolving power. As discussed in Section 3.1, a second advantage is in the reduced number of mixing products in the up-conversion process that leads only to RF combs that uniquely map to the optical region. This process significantly increases the range of parameters available for different interleaving schemes that may be implemented to reach higher orders for increased optical bandwidth coverage and higher magnifications in the RF to optical conversion to reduce the detection bandwidth. In contrast, the RF comb spectrum of the CB method for a given down-conversion scheme has limited empty spectrum because of the presence of the interleaved degenerate combs. However, we found an insufficient increase in the SNR (counts per comb tooth) for the SB method relative to the CB method despite the former having far fewer comb teeth. This uniform SNR across different methods arises in part because of the undepleted pump from the low nonlinear mixing efficiency.

**6. Conclusions**

In this paper, we have demonstrated two different methods to generate near-UV combs that span up to 120 GHz (4 cm$^{-1}$) by sum frequency generation in a single LBO crystal. Both methods rely on differential chirped pulse down conversion and interleaving of the EOM orders using a phase slip of the LO waveform to sample over a 4 kHz beat note in the near-IR.

In one method, the combined beam (CB) fiber-coupled output from the SIG and LO legs of the interferometer is focused inside the nonlinear crystal, generating an interferogram that contains a multitude of interleaved RF mixed combs, some of which retain a unique spectral map to the optical domain and some to degenerate combs that lack a unique one-to-one mapping. We compare the CB results to a more conventional separate beam (SB) method where each near-IR leg is first up-converted in crossed paths though a single crystal and then combined after sample interrogation. This crossed beam approach simplifies the phase matching conditions to a single phase-matching-angle adjustment. While the SB method always results in unique RF combs, both methods generate odd-order UV combs from mixing with the doubled carriers that pass through the EOMs. For both methods, high SNR comb spectra of the Rb isotopologues and unresolved hyperfine structure near 420 nm were obtained with RMS residual standard deviations of < 1 %. Further, because of the dual chirped-pulse down-conversion scheme, temporal magnification effects that are present only in the frequency domain are seen to distort the line shapes. However, the magnification has revealed additional high frequency structure that may be associated with the unresolved hyperfine components of the $^{87}$Rb and $^{85}$Rb isotopologues. The natural UV spectra devoid of these effects were easily recovered following the back-transformation to the time domain. From non-linear least squares fits of spectra obtained from both methods in the time domain, the best-fit frequencies and intensities of the four main-line features are in excellent agreement across four independent spectra obtained from the second- and third-order combs as evidenced by RMS residuals of < 1 %.

Finally, the advantages and disadvantages of the two methods are enumerated. The practical implementation of combining the near-UV beams of the SB method presents multiple challenges, making the CB method more favorable. Overall, the CB setup has advantages in (i) higher phase stability, (ii) higher throughput power to the detector, and (iii) the ease of alignment and use. On the other hand, the SB method (i) enables the acquisition of amplitude and phase spectra and (ii) has a far cleaner RF spectrum containing only uniquely mapped comb teeth.

**Funding**

There was no external funding for this work.

**Acknowledgments**

Special thanks to Jeeseong Hwang, Kevin Cossel, and Kimberly Briggman for their detailed comments that have improved the quality of the manuscript.

**Disclosures**

Certain commercial equipment, instruments, or materials are identified in this paper in order to specify the experimental procedure adequately. Such identification is not intended to imply recommendation or endorsement by NIST, nor is it intended to imply that the materials or equipment identified are necessarily the best available for the purpose.

**Data availability**

Data is available upon reasonable request.

## 7. References

1 Aeppli, A., Kim, K., Warfield, W., Safronova, M. S., & Ye, J. (2024). Clock with 8× 10-19 Systematic Uncertainty. Physical Review Letters, 133(2), 023401

2 Alem, O., Mhaskar, R., Jiménez-Martínez, R., Sheng, D., LeBlanc, J., Trahms, L., ... & Knappe, S. (2017). Magnetic field imaging with microfabricated optically-pumped magnetometers. Optics Express, 25(7), 7849-7858.

3 Simons, M. T., Artusio-Glimpse, A. B., Robinson, A. K., Prajapati, N., & Holloway, C. L. (2021). Rydberg atom-based sensors for radio-frequency electric field metrology, sensing, and communications. Measurement: Sensors, 18, 100273.

4 Mukai, Y., Okamoto, R., & Takeuchi, S. (2022). Quantum Fourier-transform infrared spectroscopy in the fingerprint region. Optics Express, 30(13), 22624-22636.

5 Tashima, T., Mukai, Y., Arahata, M., Oda, N., Hisamitsu, M., Tokuda, K., ... & Takeuchi, S. (2024). Ultra-broadband quantum infrared spectroscopy. Optica, 11(1), 81-87.

6 Moreau, P. A., Toninelli, E., Gregory, T., Aspden, R. S., Morris, P. A., & Padgett, M. J. (2019). Imaging Bell-type nonlocal behavior. Science Advances, 5(7), eaaw2563.

7 Okubo, S., Iwakuni, K., Inaba, H., Hosaka, K., Onae, A., Sasada, H., & Hong, F. L. (2015). Ultra-broadband dual-comb spectroscopy across 1.0–1.9 µm. Applied Physics Express, 8(8), 082402.

8 Truong, G. W., Waxman, E. M., Cossel, K. C., Baumann, E., Klose, A., Giorgetta, F. R., ... & Coddington, I. (2016). Accurate frequency referencing for fieldable dual-comb spectroscopy. Optics Express, 24(26), 30495-30504

9 Adler, F., Masłowski, P., Foltynowicz, A., Cossel, K. C., Briles, T. C., Hartl, I., & Ye, J. (2010). Mid-infrared Fourier transform spectroscopy with a broadband frequency comb. Optics Express, 18(21), 21861-21872.

10 Ideguchi, T., Poisson, A., Guelachvili, G., Picqué, N., & Hänsch, T. W. (2014). Adaptive real-time dual-comb spectroscopy. Nature Communications, 5(1), 3375.

11 Coddington, I., Newbury, N., & Swann, W. (2016). Dual-comb spectroscopy. Optica, 3(4), 414-426.

12 Majewski, W. A. (1983). A tunable, single frequency UV source for high resolution spectroscopy in the 293–330 nm range. Optics Communications, 45(3), 201-206.

13 Wilson, A. C., Ospelkaus, C., VanDevender, A. P., Mlynek, J. A., Brown, K. R., Leibfried, D., & Wineland, D. J. (2011). A 750-mW, continuous-wave, solid-state laser source at 313 nm for cooling and manipulating trapped 9 Be+ ions. Applied Physics B, 105, 741-748.

14 Freegarde, T., Coutts, J., Walz, J., Leibfried, D., & Hänsch, T. W. (1997). General analysis of type I second-harmonic generation with elliptical Gaussian beams. Journal of the Optical Society of America B, 14(8), 2010-2016.

15 Hao, Z., Wang, J., Ma, S., Mao, W., Bo, F., Gao, F., ... & Xu, J. (2017). Sum-frequency generation in on-chip lithium niobate microdisk resonators. Photonics Research, 5(6), 623-628.

16 Miller, S., Luke, K., Okawachi, Y., Cardenas, J., Gaeta, A. L., & Lipson, M. (2014). On-chip frequency comb generation at visible wavelengths via simultaneous second-and third-order optical nonlinearities. Optics Express, 22(22), 26517-26525.

17 Wang, L., Chang, L., Volet, N., Pfeiffer, M. H., Zervas, M., Guo, H., ... & Bowers, J. E. (2016). Frequency comb generation in the green using silicon nitride microresonators. Laser & Photonics Reviews, 10(4), 631-638.

18 Sun, Y., Stone, J., Lu, X., Zhou, F., Song, J., Shi, Z., & Srinivasan, K. (2024). Advancing on-chip Kerr optical parametric oscillation towards coherent applications covering the green gap. Light: Science & Applications, 13(1), 201.

19 Jiang, H., Luo, R., Liang, H., Chen, X., Chen, Y., & Lin, Q. (2017). Fast response of photorefraction in lithium niobate microresonators. Optics Letters, 42(17), 3267-3270.

20 Xue, S., Shi, Z., Ling, J., Gao, Z., Hu, Q., Zhang, K., ... & Lin, Q. (2023). Full-spectrum visible electro-optic modulator. Optica, 10(1), 125-126.

21 Martín-Mateos, P., Jerez, B., Largo-Izquierdo, P., & Acedo, P. (2018). Frequency accurate coherent electro-optic dual-comb spectroscopy in real-time. Optics Express, 26(8), 9700-9713.

22 Li, D., Ren, X., Yan, M., Fan, D., Yang, R., & Dai, H. (2021). Rapid and precise partial pressure measurement of multiple gas species with mid-infrared electro-optic dual-comb spectroscopy. Optik, 242, 167341.

23 Long, D. A., Reschovsky, B. J., LeBrun, T. W., Gorman, J. J., Hodges, J. T., Plusquellic, D. F., & Stroud, J. R. (2022). High dynamic range electro-optic dual-comb interrogation of optomechanical sensors. Optics Letters, 47(17), 4323-4326.

24 Soriano-Amat, M., Soto, M. A., Duran, V., Martins, H. F., Martin-Lopez, S., Gonzalez-Herraez, M., & Fernández-Ruiz, M. R. (2020). Common-path dual-comb spectroscopy using a single electro-optic modulator. Journal of Lightwave Technology, 38(18), 5107-5115.

25 Stroud, J. R., Long, D. A., & Plusquellic, D. F. (2024). Single-modulator, dual comb serrodyne spectroscopy. Optics Letters, 49(14), 3878-3881.

26 Han, K., Long, D. A., Bresler, S. M., Song, J., Bao, Y., Reschovsky, B. J., ... & LeBrun, T. W. (2024). Low-power, agile electro-optic frequency comb spectrometer for integrated sensors. Optica, 11(3), 392-398.

27 Wen, X., Han, Y., & Wang, J. (2016). Comparison and characterization of efficient frequency doubling at 397.5 nm with PPKTP, LBO and BiBO crystals. Laser Physics, 26(4), 045401.

28 Hickstein, D. D., Carlson, D. R., Kowligy, A., Kirchner, M., Domingue, S. R., Nader, N., ... & Diddams, S. A. (2017). High-harmonic generation in periodically poled waveguides. Optica, 4(12), 1538-1544.

29 Xu, B., Chen, Z., Hänsch, T. W., & Picqué, N. (2024). Near-ultraviolet photon-counting dual-comb spectroscopy. Nature, 627(8003), 289-294.

30 Yost, D. C., Schibli, T. R., Ye, J., Tate, J. L., Hostetter, J., Gaarde, M. B., & Schafer, K. J. (2009). Vacuum-ultraviolet frequency combs from below-threshold harmonics. Nature Physics, 5(11), 815-820.
31 Cingöz, A., Yost, D. C., Allison, T. K., Ruehl, A., Fermann, M. E., Hartl, I., & Ye, J. (2012). Direct frequency comb spectroscopy in the extreme ultraviolet. Nature, 482(7383), 68-71.
32 Kim, H., Han, S., Kim, Y. W., Kim, S., & Kim, S. W. (2017). Generation of coherent extreme-ultraviolet radiation from bulk sapphire crystal. ACS Photonics, 4(7), 1627-1632.
33 Seres, J., Seres, E., Serrat, C., Young, E. C., Speck, J. S., & Schumm, T. (2019). All-solid-state VUV frequency comb at 160 nm using high-harmonic generation in nonlinear femtosecond enhancement cavity. Optics Express, 27(5), 6618-6628.
34 Plusquellic, D. F., Wagner, G. A., Fleisher, A. J., Long, D. A., & Hodges, J. T. (2017, May). Multiheterodyne spectroscopy using multi-frequency combs. In CLEO: Applications and Technology (pp. AM1A-5). Optica Publishing Group.
35 Stroud, J. R., Simon, J. B., Wagner, G. A., & Plusquellic, D. F. (2021). Interleaved electro-optic dual comb generation to expand bandwidth and scan rate for molecular spectroscopy and dynamics studies near 1.6 µm. Optics Express, 29(21), 33155-33170.
36 Stroud, J. R., & Plusquellic, D. F. (2022). Difference-frequency chirped-pulse dual-comb generation in the THz region: Temporal magnification of the quantum dynamics of water vapor lines by> 60 000. The Journal of Chemical Physics, 156(4).
37 Stroud, J. R., & Plusquellic, D. F. (2022). Dual chirped-pulse electro-optical frequency comb method for simultaneous molecular spectroscopy and dynamics studies: formic acid in the terahertz region. Optics Letters, 47(15), 3716-3719.
38 Long, D. A., Stroud, J. R., Reschovsky, B. J., Bao, Y., Zhou, F., Bresler, S. M., ... & Gorman, J. J. (2023). High accuracy, high dynamic range optomechanical accelerometry enabled by dual comb spectroscopy. APL Photonics, 8(9).
39 Steber, A. L., Neill, J. L., Zaleski, D. P., Pate, B. H., Lesarri, A., Bird, R. G., ... & Pratt, D. W. (2011). Structural studies of biomolecules in the gas phase by chirped-pulse Fourier transform microwave spectroscopy. Faraday Discussions, 150, 227-242.
40 Brown, G. G., Dian, B. C., Douglass, K. O., Geyer, S. M., & Pate, B. H. (2006). The rotational spectrum of epifluorohydrin measured by chirped-pulse Fourier transform microwave spectroscopy. Journal of Molecular Spectroscopy, 238(2), 200-212.
41 Gerecht, E., Douglass, K. O., & Plusquellic, D. F. (2011). Chirped-pulse terahertz spectroscopy for broadband trace gas sensing. Optics Express, 19(9), 8973-8984.
42 Long, D. A., Fleisher, A. J., Plusquellic, D. F., & Hodges, J. T. (2016). Multiplexed sub-Doppler spectroscopy with an optical frequency comb. Physical Review A, 94(6), 061801.
43 Long, D. A., Bresler, S. M., Bao, Y., Reschovsky, B. J., Hodges, J. T., Lawall, J. R., ... & Gorman, J. J. (2023). Single-modulator, direct frequency comb spectroscopy via serrodyne modulation. Optics letters, 48(4), 892-895.
44 Riedle, E., Ashworth, S. H., Farrell Jr, J. T., & Nesbitt, D. J. (1994). Stabilization and precise calibration of a continuous-wave difference frequency spectrometer by use of a simple transfer cavity. Review of Scientific Instruments, 65(1), 42-48.
45 Plusquellic, D. F., Davis, S. R., & Jahanmir, F. (2001). Probing nuclear quadrupole interactions in the rotationally resolved S 1← S electronic spectrum of 2-chloronaphthalene. The Journal of chemical physics, 115(1), 225-235.
46 Hannig, S., Mielke, J., Fenske, J. A., Misera, M., Beev, N., Ospelkaus, C., & Schmidt, P. O. (2018). A highly stable monolithic enhancement cavity for second harmonic generation in the ultraviolet. Review of Scientific Instruments, 89(1).
47 Freegarde, T., Coutts, J., Walz, J., Leibfried, D., & Hänsch, T. W. (1997). General analysis of type I second-harmonic generation with elliptical Gaussian beams. Journal of the Optical Society of America B, 14(8), 2010-2016.
48 Glaser, C., Karlewski, F., Kluge, J., Grimmel, J., Kaiser, M., Günther, A., ... & Fortágh, J. (2020). Absolute frequency measurement of rubidium 5 S-6 P transitions. Physical Review A, 102(1), 012804.